\documentclass[prl,aps,showpacs,reprint]{revtex4-1}
\usepackage{amsmath}
\usepackage{amssymb}
\usepackage{graphicx}
\usepackage[bookmarks=false]{hyperref}
\usepackage{bm}
\hypersetup{colorlinks=true,citecolor=blue,linkcolor=red,urlcolor=blue,pdfstartview=FitH,bookmarksopen=true}

\usepackage[T1]{fontenc}
\usepackage{txfonts}

\DeclareMathOperator{\Tr}{Tr}
\DeclareMathOperator{\diag}{diag}
\newcommand{\bra}[1]{\langle #1\rvert}
\newcommand{\ket}[1]{\lvert #1\rangle}
\newcommand{\abs}[1]{\lvert #1\rvert}
\newcommand{\norm}[1]{\lVert #1\rVert}
\renewcommand{\i}{\mathrm{i}}
\newcommand{\e}{\mathrm{e}}
\newcommand{\I}{\mathrm{I}}
\newcommand{\tr}{\mathrm{tr}}

\begin{document}

\title{Alternative framework for quantifying coherence}
\author{Xiao-Dong Yu}
\affiliation{Department of Physics, Shandong University, Jinan 250100, China}
\author{Da-Jian Zhang}
\affiliation{Department of Physics, Shandong University, Jinan 250100, China}
\author{G. F. Xu}
\affiliation{Department of Physics, Shandong University, Jinan 250100, China}
\author{D. M. Tong}
\email{tdm@sdu.edu.cn}
\affiliation{Department of Physics, Shandong University, Jinan 250100, China}
\date{\today}

\begin{abstract}
We propose an alternative framework for quantifying coherence. The framework is based on a natural property of coherence, \textit{the additivity of coherence for subspace-independent states}, which is described by an operation-independent equality rather than operation-dependent inequalities and therefore applicable to various physical contexts. Our framework is compatible with all the known results on coherence measures but much more flexible and convenient for applications, and by using it many open questions can be resolved.
\end{abstract}
\maketitle

Quantum coherence is a fundamental feature of quantum mechanics, describing the capability of a quantum state to exhibit quantum interference phenomena. It is an essential ingredient in quantum information processing \cite{Nielsen.Chuang2000}, and plays a central role in emergent fields, such as quantum metrology \cite{Giovannetti.etal2004,Demkowicz-Dobrzanski.Maccone2014,Giovannetti.etal2011}, nanoscale thermodynamics \cite{Aberg2014,Narasimhachar.Gour2015,Cwiklinski.etal2015,Lostaglio.etal2015a,Lostaglio.etal2015b,Vazquez.etal2012,Karlstroem.etal2011}, and quantum biology \cite{Plenio.Huelga2008,Rebentrost.etal2009,Lloyd2011,Li.etal2012,Huelga.Plenio2013}. Although the theory of quantum coherence is historically well developed in quantum optics \cite{Glauber1963,Sudarshan1963,Mandel.Wolf1995}, it is only in recent years that the quantification of coherence has attracted a growing interest \cite{Aberg2006,Gour.Spekkens2008,Baumgratz.etal2014,Marvian.Spekkens2014,Levi.Mintert2014} due to the development of quantum information science.

By following the approach that has been established for entanglement resource \cite{Plenio.Virmani2007,Horodecki.etal2009}, Baumgratz \textit{et al.} proposed a seminal framework for quantifying coherence as a resource in Ref. \cite{Baumgratz.etal2014}. The framework comprises four conditions, of which the first two are based on the notions of free states and free operations in the resource theories, while the third and fourth conditions are two constraints imposed on coherence measures. Based on this framework, a number of coherence measures, such as the relative entropy of coherence, the $l_1$ norm of coherence, and the coherence of formation \cite{Aberg2006,Baumgratz.etal2014,Yuan.etal2015,Winter.Yang2016}, have been put forward. With the coherence measures, various properties of quantum coherence, such as the relations between quantum coherence and quantum correlations \cite{Streltsov.etal2015,Yao.etal2015,Xi.etal2015,Ma.etal2016,Radhakrishnan.etal2016}, the freezing phenomenon of coherence \cite{Bromley.etal2015,Yu.etal2016}, and the distillation of coherence \cite{Winter.Yang2016,Chitambar.etal2016}, were investigated. Hereafter, we refer to the framework proposed by Baumgratz \textit{et al.} as the BCP framework for simplicity.

Although the BCP framework has been widely used as an approach to coherence measures, there are arguments against the necessity of its last two conditions \cite{Streltsov.etal2015,Napoli.etal2016}, and researchers have different opinions on the definition of free operations. Besides the incoherent operations defined in the BCP framework, there have been many different suggestions on the definition of free operations, such as maximally incoherent operations \cite{Aberg2006}, translationally invariant operations \cite{Marvian.Spekkens2014}, and others \cite{Yadin.etal2016,Streltsov2015,Chitambar.Gour2016}. These arguments against the conditions and free operations imply that the frameworks for quantifying coherence are not unique. There can be other frameworks different from the BCP framework. For instance, the framework proposed by Marvian and Spekkens in Ref. \cite{Marvian.Spekkens2014}, called the MS framework for simplicity, is based on the translationally invariant operations, and it comprises only two conditions, which correspond to the first two of the BCP framework.

Possibly, based on different physical contexts, various frameworks for quantifying coherence can be constructed, and each of them may be with different conditions. The question is then: What basic conditions should be included in a well-defined framework for quantifying coherence, or is there a framework consisting of the basic conditions that can avoid the arguments against the previous conditions and are applicable to various physical contexts? In this Rapid Communication, we address this issue. We will put forward an alternative framework for quantifying coherence. It consists of three basic conditions, which are applicable to various physical contexts and can avoid the arguments against the previous frameworks. Our framework is compatible with all the known results on coherence measures but much more flexible and convenient for applications, and by using it, many open questions or arguments can be easily resolved.

To present our framework clearly, it is instructive to recapitulate some notions in resource theories, such as free states and free operations. In general resource theories, the notions of free states and free operations are, respectively, for the states that contain no resource and the operations that are unable to create the resource \cite{Coecke.etal2016,Brandao.Gour2015}. Specifically to the coherence, a free state means a quantum state with no coherence, known as an incoherent state in general, and a free operation means a special quantum operation under which coherence does not increase, known as an incoherent operation in the BCP framework. Noting that the coherence of a state is with respect to a fixed basis, known as the incoherent basis, we hereafter use $\{\ket{i},~i=0,1,\dots,d-1\}$ to denote the incoherent basis of a $d$-dimensional quantum system $\mathcal{S}$. An incoherent state can then be written as $\rho=\sum_{i}\rho_{ii}\ket{i}\bra{i}$, and a general state can be written as $\rho=\sum_{i,j}\rho_{ij}\ket{i}\bra{j}$ with coefficients $\rho_{ij}$ being the elements of the density matrix. We further use $C(\rho)$ to denote the coherence of a state $\rho$, and $\Lambda$ to denote a free operation, which can be an incoherent operation or any other operations as mentioned above. With these notions, we may now be able to develop our framework consisting of three essential conditions.

The first two of the three conditions are based on the resource theories, just like those in the previous frameworks. One of them originates from the notion of incoherent states. By definition, an incoherent state means a state with no coherence. It is natural to set the coherence of an incoherent state to zero and let the coherence of a nonincoherent state (coherent state) be positive. That is, $C(\rho)=0$ for the incoherent states and $C(\rho)>0$ for all other states. The other condition comes directly from the notion of free operations. By definition, the coherence of a state does not increase under a free operation. We then have $C(\rho)\ge C(\Lambda(\rho))$.

Our third condition is based on a characteristic of coherence itself. To develop it, we consider a special family of states, which is with the form of block-diagonal matrices, $\rho=p_1\rho_1\oplus p_2\rho_2$, i.e., $\rho=\left(\begin{smallmatrix}p_1\rho_1&0\\0& p_2\rho_2\end{smallmatrix}\right)$, where density operators $\rho_1$ and $\rho_2$ are defined on the two independent subspaces $\mathcal{S}_1$ and $\mathcal{S}_2$, respectively, and $p_1$ and $p_2$ are two possibility coefficients with $p_1+p_2=1$. Note that the coherence of a state stems from the correlations or superpositions of the incoherent basis states. Since there is no correlation between the two subspace-independent states $\rho_1$ and $\rho_2$, the coherence of $\rho$ only comes from the inside correlations of each $\rho_i$, $i=1,2$. Therefore, the coherence of $\rho$ should not be more than the total coherence of $\rho_1$ and $\rho_2$. Similarly, it should not be less than the latter either, since $\rho$ contains all the information of $\rho_1$ and $\rho_2$. By noting that $\rho$ is the mixture of $\rho_1$ and $\rho_2$ with weights $p_1$ and $p_2$, a reasonable condition can then be expressed as
\begin{equation}
  C(p_1\rho_1\oplus p_2\rho_2)=p_1C(\rho_1)+p_2C(\rho_2).
  \label{condition3}
\end{equation}
Expression \eqref{condition3} is derived from the characteristic of quantum coherence. It is a display of a basic property of coherence, to which we refer as \textit{the additivity of coherence for subspace-independent states}.

The above three conditions form a general framework for quantifying coherence. If we further specify incoherent operations for free operations, our framework can be expressed as follows. A functional $C$ can be taken as a coherence measure if it satisfies the following three conditions:
  \\(C1) $C(\rho)\ge 0$ for all states, and $C(\rho)=0$ if and only if $\rho$ are incoherent states;
  \\(C2) $C(\rho)\ge C(\Lambda(\rho))$ if $\Lambda$ is an incoherent operation; and
  \\(C3) $C(p_1\rho_1\oplus p_2\rho_2)=p_1C(\rho_1)+p_2C(\rho_2)$ for block-diagonal states $\rho$ in the incoherent basis.

The third condition in our framework is described by only an operation-independent equality rather than operation-dependent inequalities. This is the key to our framework. Many advantages of the framework, such as the compatibility, flexibility and applicability, originate from this simple equality. The following discussions will further elucidate the merits of our framework.

First, we show that the additivity of coherence for subspace-independent states, i.e., our third condition, is fulfilled by all the coherence measures based on the BCP framework. To make the proof clear, we recall the BCP framework. A functional $C$ can be taken as a coherence measure if it satisfies the following four conditions:
  \\(B1) $C(\rho)\ge 0$ for all states, and $C(\rho)=0$ if and only if $\rho$ is an incoherent state;
  \\(B2) $C(\rho)\ge C(\Lambda(\rho))$ if $\Lambda$ is an incoherent operation, i.e., a completely positive trace-preserving (CPTP) map $\Lambda(\rho)=\sum_nK_n\rho K_n^\dagger$ with the Kraus operators $K_n$ satisfying $K_n\mathcal{I}K_n^\dagger\subset\mathcal{I}$, where $\mathcal{I}$ is the set of incoherent states \footnote{An incoherent operation can be a CPTP map with $K_n$ being $(M\times N)$-dimensipnal matrices, which satisfy $K_n\mathcal{I}_{N}K_n\subset\mathcal{I}_{M}$, not confined to $M=N$. Here $\mathcal{I}_{N}$ ($\mathcal{I}_{M}$) is the set of incoherent states of the $N$($M$)-dimensional space.};
  \\(B3) $C(\rho)\ge\sum_np_nC(\rho_n)$, where $p_n=\Tr(K_n\rho K_n^\dagger)$, $\rho_n=K_n\rho K_n^\dagger/p_n$, and $K_n$ are the Kraus operators of an incoherent CPTP map $\Lambda(\rho)=\sum_nK_n\rho K_n^\dagger$; and
  \\(B4) $\sum_np_nC(\rho_n)\ge C(\sum_np_n\rho_n)$ for any set of states $\{\rho_n\}$ and any probability distribution $\{p_n\}$.

We aim to show that any $C$ satisfying the conditions (B1)--(B4) will necessarily satisfy $C(p_1\rho_1\oplus p_2\rho_2)=p_1C(\rho_1)+p_2C(\rho_2)$. To this end, we consider an incoherent CPTP map, $\Lambda(\cdot)=P_1 \cdot P_1^\dagger+P_2\cdot P_2^\dagger$, where $P_1=\ket{0}\bra{0}+\cdots+\ket{N_1-1}\bra{N_1-1}$ and $P_2=\ket{N_1}\bra{N_1}+\cdots+\ket{N_1+N_2-1}\bra{N_1+N_2-1}$ are projectors onto $\mathcal{S}_1$ and $\mathcal{S}_2$, with $N_1$ and $N_2$ being the dimensions of $\mathcal{S}_1$ and $\mathcal{S}_2$, respectively. It is easy to verify $P_n\mathcal{I}P_n^\dagger\subset\mathcal{I}$. According to (B3), since $\Tr[P_1(p_1\rho_1\oplus p_2\rho_2)P_1^\dagger]=p_1$, $\Tr[P_2(p_1\rho_1\oplus p_2\rho_2)P_2^\dagger]=p_2$, $P_1\left(p_1\rho_1\oplus p_2\rho_2\right)P_1^\dagger/p_1=\rho_1\oplus 0$, and $P_2\left(p_1\rho_1\oplus p_2\rho_2\right)P_2^\dagger/p_2=0\oplus\rho_2$, we obtain
\begin{equation}
C(p_1\rho_1\oplus p_2\rho_2)\ge p_1C(\rho_1\oplus 0)+p_2C(0\oplus\rho_2).
\label{eq:linearity_le}
\end{equation}
On the other hand, according to (B4), since $p_1\rho_1\oplus p_2\rho_2=p_1\left(\rho_1\oplus 0\right)+p_2\left(0\oplus\rho_2\right)$, we have
\begin{equation}
  C(p_1\rho_1\oplus p_2\rho_2)\le p_1C(\rho_1\oplus 0)+p_2C(0\oplus\rho_2).
  \label{eq:linearity_ge}
\end{equation}
However, $C(\rho)$, as a valid coherence measure based on the BCP framework, must satisfy both Eqs. \eqref{eq:linearity_le} and \eqref{eq:linearity_ge}. This results in an equality,
\begin{equation}
  C(p_1\rho_1\oplus p_2\rho_2)=p_1C(\rho_1\oplus 0)+p_2C(0\oplus\rho_2).
  \label{eq:linearity_approx}
\end{equation}

To obtain (C3), we need to prove $C(\rho_1\oplus 0)=C(\rho_1)$. To this end, we consider two incoherent CPTP maps, $\Lambda^a(\cdot)=K_0^a\cdot K_0^{a\dagger}$ with $\bra{i}K_0^a\ket{j}=\delta_{ij}$ and $\Lambda^b(\cdot)=\sum_{n=0}^{\lceil\frac{N_2}{N_1}\rceil}K_n^b\cdot K_n^{b\dagger}$ with $\bra{j}K_n^b\ket{i}=\delta_{i,j+nN_1}$, where $0\le i\le N_1+N_2-1$, $0\le j\le N_1-1$, and $\lceil\frac{N_2}{N_1}\rceil$ is the smallest integer greater than or equal to $\frac{N_2}{N_1}$. It is easy to verify that both $\Lambda^a$ and $\Lambda^b$ are incoherent CPTP maps, and there are $\Lambda^a(\rho_1)=\rho_1\oplus 0$ and $\Lambda^b(\rho_1\oplus 0)=\rho_1$. Thus, according to condition (B2), we should have $C(\rho_1)\le C(\rho_1\oplus 0)\le C(\rho_1)$, which results in $C(\rho_1\oplus 0)=C(\rho_1)$. Similarly, we can prove $C(0\oplus\rho_2)=C(\rho_2)$. By substituting these relations into Eq.(\ref{eq:linearity_approx}), we finally obtain $C(p_1\rho_1\oplus p_2\rho_2)=p_1C(\rho_1)+p_2C(\rho_2)$.

Second, we show that the BCP framework can be derived from our framework. That is, each of the four conditions (B1)--(B4) can be derived from (C1), (C2), and (C3). To prove it, we only need to derive (B3) and (B4) from conditions (C1), (C2), and (C3), since (B1) and (B2) are just corresponding to (C1) and (C2).

We first derive (B3). To this end, we introduce an auxiliary system $\mathcal{A}$ of dimension $N$, of which the incoherent basis is denoted as $\{\ket{n},~0\le n\le N-1\}$. The auxiliary system $\mathcal{A}$ and the system $\mathcal{S}$ form a combined system $\mathcal{AS}$, of which the incoherent basis is $\{\ket{n}\otimes\ket{i}\}$. We suppose that the whole system is initially in the state,
\begin{equation}
  \rho^{\mathcal{AS}}=\ket{0}\bra{0}\otimes\rho,
  \label{rhoas}
\end{equation}
and undergoes an incoherent CPTP map,
\begin{equation}
  \Lambda^{\mathcal{AS}}(\rho^{\mathcal{AS}})=\sum_{n=0}^{N-1}(U_n\otimes K_n)\rho^{\mathcal{AS}}(U_n\otimes K_n)^\dagger,
  \label{lambdaas}
\end{equation}
where $U_n\otimes K_n$ are the Kraus operators of $\Lambda^{\mathcal{AS}}$ with $U_n=\sum_{k=0}^{N-1}\ket{k+n\pmod{N}}\bra{k}$, and $K_n$ are the Kraus operators of the incoherent CPTP map $\Lambda$. It is easy to verify that $(U_n\otimes K_n)\mathcal{I}(U_n\otimes K_n)^\dagger\subset\mathcal{I}$. Note that $\Lambda$ has been defined as an incoherent CPTP map. Substituting Eq. \eqref{rhoas} into Eq. \eqref{lambdaas}, we have
\begin{equation}
  \Lambda^{\mathcal{AS}}(\rho^{\mathcal{AS}})=\sum_{n=0}^{N-1}p_n\ket{n}\bra{n}\otimes\rho_n,
  \label{lambdaas2}
\end{equation}
where $p_n=\Tr(K_n\rho K_n^\dagger)$ and $\rho_n=K_n\rho K_n^\dagger/p_n$.

According to condition (C3), we have
\begin{equation}
C\left(\rho^{\mathcal{AS}}\right)=C\left(\ket{0}\bra{0}\otimes\rho\right)=C\left(\rho\oplus 0\right)=C\left(\rho\right),
  \label{crho1}
\end{equation}
and
\begin{equation}
  \begin{aligned}
    C\left(\Lambda^{\mathcal{AS}}(\rho^{\mathcal{AS}})\right)&=C\left(\sum_{n=0}^{N-1}p_n\ket{n}\bra{n}\otimes\rho_n\right)\\
    &=C\left(p_0\rho_0\oplus p_1\rho_1\oplus\cdots\oplus p_{N-1}\rho_{N-1}\right)\\
    &=\sum_{n=0}^{N-1}p_nC(\rho_n).
  \end{aligned}
  \label{crho22}
\end{equation}
According to condition (C2), from Eqs. \eqref{crho1} and \eqref{crho22} we immediately derive
\begin{equation}
  C(\rho)\ge\sum_np_nC(\rho_n),
  \label{ge}
\end{equation}
i.e., the condition (B3).

By the way, one may find that the first equality in Eq. \eqref{crho22} immediately leads to the condition $C(\rho)\ge C\left(\sum_{n=0}^{N-1}p_n\ket{n}\bra{n}\otimes\rho_n\right)$, i.e., the classical flag monotonicity \cite{Baumgratz.etal2014}. It implies that the classical flag monotonicity, which was proved only for the relative entropy of coherence and $l_1$ norm of coherence, is actually valid for all coherence measures satisfying our third condition and therefore for all the coherence measures based on the BCP framework.

We now derive (B4). We again consider the combined system comprising the auxiliary system $\mathcal{A}$ and the system $\mathcal{S}$, as stated above. Now, we suppose the whole system is initially in the state,
\begin{equation}
  \rho^{\mathcal{AS}}=\sum_{n=0}^{N-1}p_n\ket{n}\bra{n}\otimes\rho_n,
  \label{rhoas2}
\end{equation}
where $\{\rho_n,~0\le n\le N-1\}$ is a set of states and $\{p_n,~0\le n\le N-1\}$ is a probability distribution, and undergoes an incoherent CPTP map,
\begin{equation}
  \Lambda^{\mathcal{AS}}(\rho^{\mathcal{AS}})=\sum_{n=0}^{N-1}(\ket{0}\bra{n}\otimes\I)\rho^{\mathcal{AS}}(\ket{0}\bra{n}\otimes\I)^\dagger,
  \label{lambdaas3}
\end{equation}
where $\ket{0}\bra{n}\otimes\I$ are the Kraus operators of $\Lambda^{\mathcal{AS}}$, satisfying $(\ket{0}\bra{n}\otimes\I)\mathcal{I}(\ket{0}\bra{n}\otimes\I)^\dagger\subset\mathcal{I}$. Substituting Eq. \eqref{rhoas2} into Eq. \eqref{lambdaas3}, we have
\begin{equation}
  \Lambda^{\mathcal{AS}}(\rho^{\mathcal{AS}})=\ket{0}\bra{0}\otimes\sum_{n=0}^{N-1}p_n\rho_n.
  \label{lambdaas4}
\end{equation}

According to condition (C3), we have
\begin{equation}
  C\left(\rho^{\mathcal{AS}}\right)=C\left(\sum_{n=0}^{N-1}p_n\ket{n}\bra{n}\otimes\rho_n\right)=\sum_{n=0}^{N-1}p_nC(\rho_n),
  \label{crho3}
\end{equation}
and
\begin{equation}
  C\left(\Lambda^{\mathcal{AS}}(\rho^{\mathcal{AS}})\right)=C\left(\ket{0}\bra{0}\otimes\sum_{n=0}^{N-1}p_n\rho_n\right)=C\left(\sum_{n=0}^{N-1}p_n\rho_n\right).
  \label{crho4}
\end{equation}
According to condition (C2), from Eqs. \eqref{crho3} and \eqref{crho4} we immediately derive
\begin{equation}
  \sum_{n=0}^{N-1}p_nC(\rho_n)\ge C\left(\sum_{n=0}^{N-1}p_n\rho_n\right),
  \label{eq:convex}
\end{equation}
i.e., condition (B4).

Third, our framework is efficient and convenient for applications, and it can help to resolve some open questions and arguments. For instance, our framework can help to resolve the argument about the necessity of the last two conditions in the BCP framework. As mentioned above, the two conditions (B3) and (B4) have been suspected of necessity. Here, our discussion shows that these conditions can be derived from the natural property of coherence described by Eq. \eqref{condition3}, and therefore they are reasonable requirements in the BCP framework. Furthermore, our framework can simplify the calculations in examining whether a functional $C$ is qualified as a coherence measure. Generally speaking, it is relatively easier to examine whether a candidate of coherence measure satisfies (C3) than to examine whether it satisfies (B3) and (B4), since (C3) is only an equality and does not involve operations. For example, the proof of the relative entropy of coherence can be significantly simplified by using our framework, since condition (C3) follows directly from the well-known relation for entropy $S(p_1\rho_1\oplus p_2\rho_2)=H(p_1,p_2)+p_1S(\rho_1)+p_2S(\rho_2)$, where $S$ is von Neumann entropy and $H$ is Shannon entropy \cite{Nielsen.Chuang2000}. In the following, we would like to give one more example, i.e., the trace norm of coherence, to further show the efficiency of our framework.

The trace norm of coherence is defined as
\begin{equation}
  C_{\tr}(\rho)=\min_{\delta\in\mathcal{I}}\norm{\rho-\delta}_{\tr},
  \label{eq:C_tr}
\end{equation}
where $\norm{\rho-\delta}_{\tr}=\Tr\abs{\rho-\delta}$ is the trace norm between the state $\rho$ and the incoherent states $\delta$ \cite{Baumgratz.etal2014}. $C_{\tr}$ has been expected to be a coherence measure, but it is quite difficult to prove it to satisfy all four conditions in the BCP framework. So far, whether the trace norm of coherence is a legitimate coherence measure is still an open question. Previous works have proved that $C_{\tr}$ satisfies (B1), (B2), and (B4) \cite{Baumgratz.etal2014,Bromley.etal2015}. Recently, it was further proved that (B3) is fulfilled at least for qubit and $X$ states \cite{Shao.etal2015,Rana.etal2016}. Yet, it remains unknown whether (B3) is fulfilled for all other states. By using our framework, the open question is resolved, since $C_{\tr}$ does not satisfy our third condition. To illustrate this, we need first to prove
\begin{equation}
C_{\tr}(\ket{\Psi_d}\bra{\Psi_d})=\min_{\delta\in\mathcal{I}}\norm{\ket{\Psi_d}\bra{\Psi_d}-\delta}_{\tr}=\frac{2(d-1)}{d},
\label{Cpsid}
\end{equation}
where $\ket{\Psi_d}=\frac{1}{\sqrt{d}}\sum_{n=0}^{d-1}\ket{n}$, $d\ge 1$. For this, we let $U_n=\sum_{k=0}^{d-1}\ket{k+n\pmod d}\bra{k}$, $n=0,1,\dots,d-1$. By using the relations $\norm{UAU^\dagger}_{\tr}=\norm{A}_{\tr}$ and $\norm{A}_{\tr}+\norm{B}_{\tr}\ge\norm{A+B}_{\tr}$, which are valid for all the same dimensional matrices $A,B$ and unitary operators $U$ \cite{Bhatia1997}, we can obtain $\norm{\ket{\Psi_d}\bra{\Psi_d}-\delta}_{\tr}=\frac{1}{d}\sum_{n=0}^{d-1}\norm{U_n(\ket{\Psi_d}\bra{\Psi_d}-\delta) U_n^\dagger}_{\tr}\ge\frac{1}{d}\norm{\sum_{n=0}^{d-1}U_n(\ket{\Psi_d}\bra{\Psi_d}-\delta) U_n^\dagger}_{\tr}$. Noting that $U_n\ket{\Psi_d}=\ket{\Psi_d}$ and $\sum_{n=0}^{d-1}U_n\delta U_n^\dagger=\I_d$, we further obtain $\norm{\ket{\Psi_d}\bra{\Psi_d}-\delta}_{\tr}\ge\norm{\ket{\Psi_d}\bra{\Psi_d}-\frac{1}{d}\I_d}_{\tr}$. This inequality necessarily leads to $\min_{\delta\in\mathcal{I}}\norm{\ket{\Psi_d}\bra{\Psi_d}-\delta}_{\tr}=\norm{\ket{\Psi_d}\bra{\Psi_d}-\frac{1}{d}\I_d}_{\tr}$, which further gives the result in Eq. \eqref{Cpsid}. We then consider a special state, $\rho=\frac{1}{2}\rho_1\oplus\frac{1}{2}\rho_2$, with $\rho_1=\frac{1}{2}(\ket{0}+\ket{1})(\bra{0}+\bra{1})$ and $\rho_2=\frac{1}{3}(\ket{2}+\ket{3}+\ket{4})(\bra{2}+\bra{3}+\bra{4})$. By definition, there is $C_{\tr}(\rho)=\min_{\delta\in\mathcal{I}}\norm{\rho-\delta}_{\tr}\le\norm{\rho-\delta_0}_{\tr}=1$, where $\delta_0=\diag(\frac{1}{2},\frac{1}{2},0,0,0)$. On the other hand, from Eq. \eqref{Cpsid}, we have $C_{\tr}(\rho_1)=1$, $C_{\tr}(\rho_2)=\frac{4}{3}$, and hence $\frac{1}{2}C_{\tr}(\rho_1)+\frac{1}{2}C_{\tr}(\rho_2)=\frac{7}{6}$, which shows that
\begin{equation}
  C_{\tr}(\frac{1}{2}\rho_1\oplus\frac{1}{2}\rho_2)\neq\frac{1}{2}C_{\tr}(\rho_1)+\frac{1}{2}C_{\tr}(\rho_2).
  \label{condition12}
\end{equation}
Therefore, the trace norm of coherence is not a legitimate coherence measure, and it must violate (B3) too.

In passing, we would like to add that the modified trace norm of coherence, $C'_{\tr}(\rho)=\min_{\lambda\ge 0,\delta\in\mathcal{I}}\norm{\rho-\lambda\delta}_{\tr}$, can be proved to satisfy (C1), (C2), and (C3), and therefore provides a legitimate coherence measure \footnote{Condition (C2), i.e., $C'_{tr}(\rho)\ge C'_{tr}(\Lambda(\rho))$, immediately follows from the fact that $\norm{A}_{\tr}\ge\norm{\Lambda(A)}_{\tr}$ for all Hermite matrices $A$ and CPTP maps $\Lambda$ \cite{Rivas.etal2014}. To prove (C3), we rewrite $\lambda\delta$ as $\lambda_1\delta_1\oplus\lambda_2\delta_2$, and then we can derive the following relations, $C'_{\tr}(p_1\rho_1\oplus p_2\rho_2)=\min_{\lambda_1\ge 0,\lambda_2\ge 0,\delta_1\in\mathcal{I},\delta_2\in\mathcal{I}}\norm{p_1\rho_1\oplus p_2\rho_2-\lambda_1\delta_1\oplus\lambda_2\delta_2}_{\tr}=\min_{\lambda_1\ge 0,\delta_1\in\mathcal{I}}\norm{p_1\rho_1-\lambda_1\delta_1}_{\tr}+\min_{\lambda_2\ge 0,\delta_2\in\mathcal{I}}\norm{p_2\rho_2-\lambda_2\delta_2}_{\tr}=p_1\min_{\lambda'_1\ge 0,\delta_1\in\mathcal{I}}\norm{\rho_1-\lambda'_1\delta_1}_{\tr}+p_2\min_{\lambda'_2\ge 0,\delta_2\in\mathcal{I}}\norm{\rho_2-\lambda'_2\delta_2}_{\tr}=p_1C'_{\tr}(\rho_1)+p_2C'_{\tr}(\rho_2)$.}. However, it is quite difficult to prove this result without using our framework.

All the above discussions show that our framework, compared with the seminal BCP framework, has many interesting features. (1) An operation-independent equality in our framework takes the place of both an operation-dependent inequality and an operation-independent inequality in the BCP framework. This makes our framework simple in form and convenient for applications. (2) Our third condition is fulfilled by all the coherence measures based on the BCP framework, and the BCP framework can be derived from our framework. This compatibility can greatly simplify many calculations by using our framework instead of the BCP framework. (3) Our framework can help to resolve some open questions or arguments about quantifying coherence. For instance, the open question whether the trace norm of coherence is a legitimate coherence measure is immediately resolved by using our framework.

Before concluding, we would like to stress that our framework, as an approach for quantifying coherence, is generally applicable to various physical contexts. Although we have used the notion of incoherent operations in the expression of our framework in order to compare with the BCP framework, our framework is still valid if the incoherent operations are replaced by any other free operations as needed. For instance, the incoherent operations can be replaced by the translationally invariant operations defined in Ref. \cite{Marvian.Spekkens2014}, which leads to another expression of our framework. Note that the translationally invariant operations, developed from the resource theories of asymmetry \cite{Gour.Spekkens2008,Vaccaro.etal2008,Gour.etal2009,Toloui.etal2011,Skotiniotis.Gour2012,Marvian.Spekkens2013}, are described with the help of a fixed observable $H$, of which the eigenstates just correspond to the fixed basis in our framework. By definition, a functional $C_H$ can be taken as a coherence measure relative to a fixed observable $H$, if it satisfies the following two conditions:
\\(M1) $C_H(\rho)\ge 0$, and $C_H(\rho)=0$ if and only if $\rho$ is a translationally invariant state, i.e., satisfying $\e^{-\i Ht}\rho\e^{\i Ht}=\rho$; and
\\(M2) $C_H(\rho)\ge C_H(\Lambda(\rho))$ if $\Lambda$ is a translationally invariant operation, i.e., a CPTP map satisfying $\e^{-\i Ht}\Lambda(\rho)\e^{\i Ht}=\Lambda(\e^{-\i Ht}\rho\e^{\i Ht})$ for all states $\rho$.
\\Comparing the MS framework and our framework, one may find that (M1) and (M2) are just equivalent to (C1) and (C2) if the incoherent operations in our framework are replaced by the translationally invariant operations. However, there is one more condition in our framework. In the language of a fixed observable $H$, an equivalent expression of Eq. \eqref{condition3} can be rewritten as
\\(M3) $C_{H_1\oplus H_2}(p_1\rho_1\oplus p_2\rho_2)=p_1C_{H_1}(\rho_1)+p_2C_{H_2}(\rho_2)$,
\\where $H_i$ represent the components of observable $H$ in the subspace $\mathcal{S}_i$ on which the density operators $\rho_i$ are defined. Such expression of our framework is applicable to the translationally invariant operations. It is interesting to note that all the known coherence measures based on the MS framework, such as Dyson-Wigner-Yanase skew information \cite{Marvian.Spekkens2014,Girolami2014}, the trace norm of commutator \cite{Marvian.Spekkens2014}, and the quantum Fisher information \cite{Yadin.Vedral2016}, fulfill our third condition. The coherence measures fulfilling our third condition automatically satisfy the monotonicity of coherence under selective measurements on average and the nonincreasing of coherence under mixing of states, while the coherence measures fulfilling only the MS framework but not the third condition cannot have these features.

In conclusion, we have put forward a property of coherence, called \textit{the additivity of coherence for subspace-independent states}, which is applicable to various physical contexts, and based on it, an alternative framework for quantifying coherence is constructed. Our framework, consisting of three basic conditions, is compatible with all the known results on coherence measures but much more flexible and convenient for applications, and it can significantly improve the theories of quantifying coherence.

Our finding leads to a much simpler and more practical expression of the seminal BCP framework if the incoherent operations are specified for free operations. Many open questions, which have been proved difficult to resolve by using the previous frameworks, can be resolved by using our framework.

\begin{acknowledgments}
This work was supported by National Natural Science Foundation of China through Grant No. 11575101 and the National Basic Research Program of China through Grant No. 2015CB921004. D.J.Z. acknowledges support from the China Postdoctoral Science Foundation through Grant No. 2016M592173. G.F.X. acknowledges the support from National Natural Science Foundation of China through Grant No. 11547245. D.M.T. acknowledges support from the Taishan Scholarship Project of Shandong Province.
\end{acknowledgments}

% \bibliography{coherence}

\end{document}